# Solitary Shock Waves and Adiabatic Phase Transition in Lipid Interfaces and Nerves.

Shamit Shrivastava, Kevin Heeyong Kang, Matthias F. Schneider*

Department of Mechanical Engineering, Boston University, Boston MA 02215

**Abstract**

This study shows that the stability of solitary waves excited in a lipid monolayer near a phase boundary requires positive curvature of the adiabats, a known necessary condition in shock compression science. It is further shown that the condition results in a threshold for excitation, saturation of the wave's amplitude and the splitting of the wave at the phase boundaries. Splitting in particular confirms that a hydrated lipid interface can undergo condensation on adiabatic heating thus showing retrograde behavior. Finally, using the new theoretical insights and state dependence of conduction velocity in nerves, the curvature of the adiabatic state diagram is shown to be closely tied to the thermodynamic blockage of nerve pulse propagation.

**Introduction**

Dense networks of hydrated membrane interfaces populate cellular environments [1]. The elastic properties of these quasi-2D systems have been a subject of extensive research [2]. However, most studies usually consider quasi-static processes [3] and/or small displacements [4]. We have recently demonstrated that the elastic properties of such interfaces support the propagation of acoustic pulses, which has implications for cell communication from single cells to action potentials in the nervous system [5,6].

Particular interest arises from the observation that the cellular machinery spends considerable resources in fine tuning the thermodynamic (TD) state diagrams of biological membranes, usually adapting in the vicinity of a phase transition, where the state diagram exhibits clear nonlinearities [7,8]. For dynamic processes the curvature of the adiabatic state diagram is of fundamental importance [9,10]. Indeed, a non-linearity in the elastic properties of the plasma membrane is believed to be crucial for the phenomenon of nerve pulse propagation [11–13] which is known to be adiabatic [14]. Therefore an understanding of the adiabatic state diagrams of lipid interfaces near similar nonlinearities is crucial for an improved thermodynamic understanding of nerve pulse propagation. Interestingly, spontaneous mechanical perturbations have been observed in a variety of biological systems [15] and are found to propagate along with nerve pulse propagation as well, analogous to sound waves [16,17]. In a recent study [6] we showed that thermodynamically coupled perturbations (electrical-optical-mechanical) during 2D sound waves in a lipid monolayer, a simple model system for plasma membrane, are strikingly similar to those observed during nerve pulse propagation. Based on experimentally determined sound velocities $c(\pi)$, it appeared that solitary waves only exist for a positive curvature $\left(\frac{\partial^2 a}{\partial \pi^2}\right)$ of the state diagram ($\pi \: and \: a$ represent the lateral pressure and specific area at the lipid interface). This results in a threshold for excitation if the initial equilibrium state is within a regime of negative curvature and changes to positive during excitation/propagation

Here we investigate the velocity of propagation of such waves as a function of amplitude and show that locally the condition $\left(\frac{\partial^2 a}{\partial \pi^2}\right)_S > 0$ (S representing the interfacial entropy) is preserved for the 2D sound waves at the liquid expanded/ liquid condensed phase boundary, even though the isothermal compression $\left(\frac{\partial^2 a}{\partial \pi^2}\right)_T < 0$. Furthermore we show that the maximum amplitude saturates at a value that is significantly less than expected from isothermal compression. The evolution of these waves over distance shows splitting into a non-dispersive forerunner wave and a slower dispersive wave. These results are in accordance with classical shock theory where the condition $\left(\frac{\partial^2 v}{\partial P^2}\right)_S > 0$ [9,18] or $\left(\frac{\partial^2 P}{\partial v^2}\right)_S > 0$ [10] is associated with the existence of compression shocks and its violation is associated with rarefaction shock waves, where $P$ and $v$ are pressure and specific volume. Finally we discuss the implications of the curvature of the state diagram of the nerve membrane and its possible relation to reversible thermodynamic blockage of nerve pulse propagation.

A lipid monolayer easily self assembles by adding a lipid and fluorophore mixture at the air/water interface of a Langmuir trough. Lipid monolayers are not only accessible and robust, but their molecular composition and thermodynamic state and hence the mechanical properties can also be precisely controlled, monitored, and characterized [19]. Thus they provide an excellent platform to study 2D interfacial sound waves, both experimentally and theoretically [5,6]. The opto-mechanical setup has been described in detail elsewhere [6,20]. A cantilever excites the monolayer (containing the lipids (DPPC), the donor (NBD-PE) and acceptor dye molecules (Texas Red-DHPE) (100:1:1)) longitudinally with a piezo controlled deflection producing 2D sound waves. A microscope records these sound waves by observing the ratiometric Forster resonance energy transfer (FRET), simultaneously at two wavelengths (535 and 605nm) between a pair of lipid-conjugated fluorophores, using $\frac{\Delta\theta}{\theta} = \frac{\Delta I_{535}}{I_{535}} - \frac{\Delta I_{605}}{I_{605}}$. The distance between the cantilever and the objective can be controlled by a screw meter.

## Results and Discussion

### Threshold, Saturation, Velocity and Pulse Width

For the linear case (infinitesimal amplitude), the velocity of sound is related to state variables according to $c^2 = \frac{1}{\rho k_S} = \left(\frac{\partial \pi}{\partial \rho}\right)_S$, where $\pi$ is the lateral pressure, $\rho$ (kg/m$^2$) is the density of the quasi-2D interfacial region and $k_S$ is the isentropic compressibility of the interface [5]. For a nonlinear system, such as a lipid monolayer near a phase transition, $k_S$ strongly depends on the density change $\Delta\rho$ and can undergo significant changes within a single pulse $\Delta\rho(t)$. Therefore the velocity varies within a single pulse resulting in evolving pulse shapes [6,12]. Figure 1(a) shows this behavior for different pulses as solitary waves of different amplitudes, obtained by varying the stimulus (i.e. the mechanical impulse from the piezo device) and measured via FRET, arrive at different times for a given mean equilibrium state ($\pi = \frac{4.3 mN}{m}$ and $T = 293.15 K$) (Fig. 1b). Note that compression amplitudes ($\Delta\rho/\rho_0$) - with $\frac{\Delta\rho}{\rho_0} = -\frac{\Delta a}{a}$ - can be estimated from variations in FRET parameter $\Delta\theta/\theta$ using the characteristic curve $(a \leftrightarrow \theta)_T$ obtained during isothermal compression (Fig 1b and c) [6].

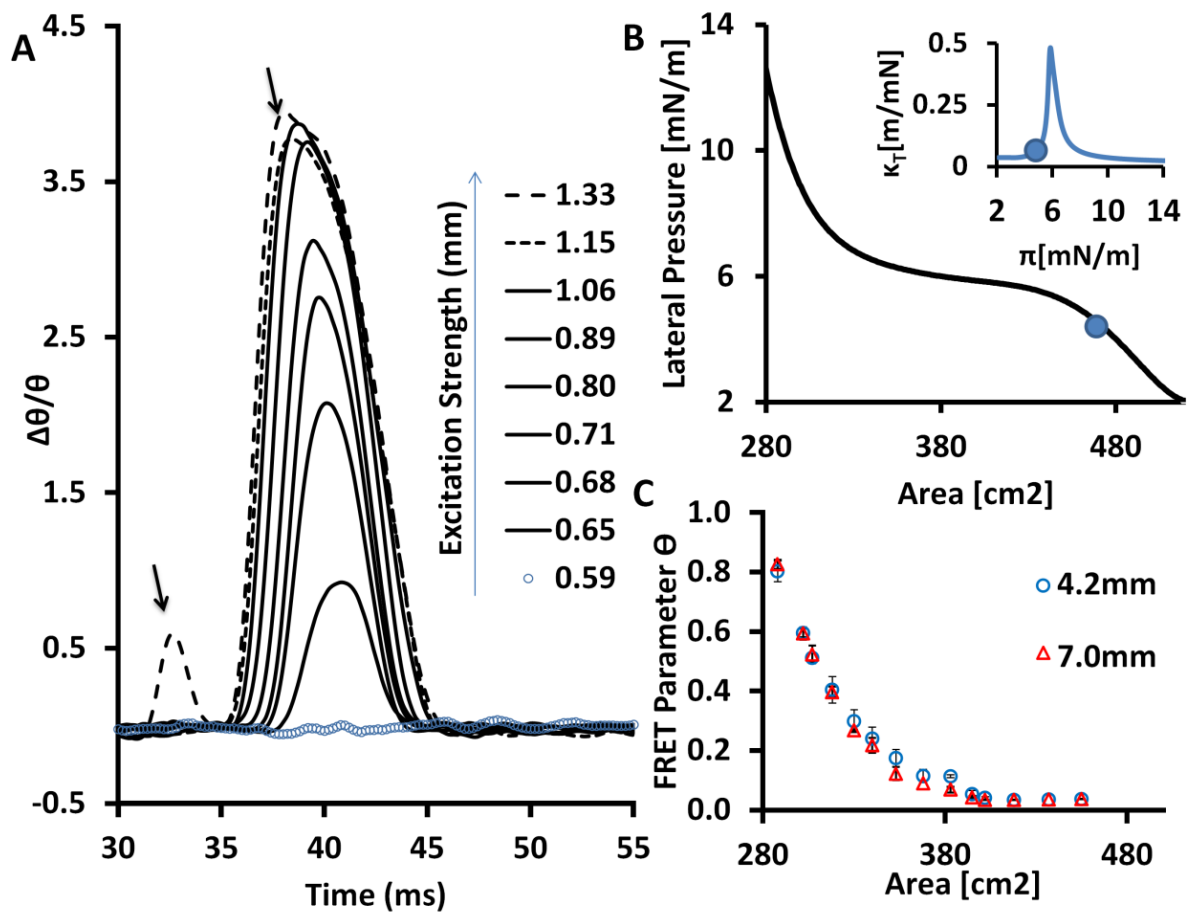

*Figure 1 (Color Online) Dependence on excitation strength and amplitude (color online). (a) Measured pulse shapes at 0.84cm from the excitation blade as a function of blade amplitude (mm). The pulses are excited at t=0. The fixed equilibrium state ($\pi = \frac{4.3mN}{m}$ and $T = 293.15K$) is indicated by the dot on (b) the isothermal state diagram and the isothermal compressibility plot in the inset. (c) The characteristic isothermal curve relating FRET parameter and surface area $(a \leftrightarrow \theta)_T$ obtained during quasi-static compression at two different spots along the propagation path. The arrows in (a) indicate the splitting of the pulse on increasing the excitation strength beyond saturation limit.*

Fluorescent probes that depend on dipole reorientation for sensing voltage changes (as observed here in lipid monolayer [6,20]), have been shown to report them without discernible time lag during nerve pulse propagation [21–23]; which indicates that as far as our overall goal of understanding thermodynamics of nerve-pulse propagation is concerned the isothermal opto-mechanical coupling is a reasonable approximation during the observed pulses as well. This allows immediate extraction of three key relations from these experiments; (i) The response ($\Delta\rho/\rho_0$) is highly nonlinear with a clear threshold and an asymptotic saturation of amplitude as a function of excitation strength (Fig 2a). (ii) The velocity calculated from the time of arrival varies linearly with relative compression ($\Delta\rho/\rho_0$) upto $\Delta\rho/\rho_0$=0.15 (Fig. 2b), which coincides with the beginning of the saturation (indicated by dashed lines Fig. 2a and b) of the nonlinear response curve. (iii) The half-width of a pulse as a function of relative compression also follows the exact same trend as velocity and diverges from a linear dependence near maximum amplitude (Fig 2b). Notably, the observed saturation of $(\Delta\rho/\rho_0)_{max}$=0.15 is approximately 20% of the value expected from the relative compression during a quasi-static phase transition ($\frac{\Delta\rho}{\rho_0} = -\frac{\Delta a}{a} = 0.75$) (Fig 1b). Further increase in the excitation strength results into splitting of the pulse (indicated by arrows) as will be discussed below. All these factors clearly indicate that a fundamental understanding of these processes has to be tied to the dynamic properties of the system.

**Positive curvature of the Adiabatic State Diagram**

It is useful to write the curvature $\left(\frac{\partial^2 a}{\partial \pi^2}\right)_S$ in its non-dimensional form $\Gamma = \frac{c^4}{2a^3}\left(\frac{\partial^2 a}{\partial \pi^2}\right)_S$, because then it is directly related to the well-known acoustic parameter of nonlinearity B/A as $\Gamma = \frac{B}{2A} + 1$. Within a thermodynamic treatment of acoustics, A and B are directly related to the coefficients of the isentropic Taylor expansion for the lateral pressure $\pi(\rho, S)$ [24];

$$\pi_S = \pi_0 + A\left(\frac{\Delta\rho}{\rho_0}\right) + \frac{B}{2}\left(\frac{\Delta\rho}{\rho_0}\right)^2 + \cdots \qquad (1)$$

with

$$A = \rho_0 \left(\frac{\partial \pi}{\partial \rho}\right)_S = \rho_0 c_0^2; B = \frac{\rho_0^2}{2!}\left(\frac{\partial^2 \pi}{\partial \rho^2}\right)_S. \tag{2}$$

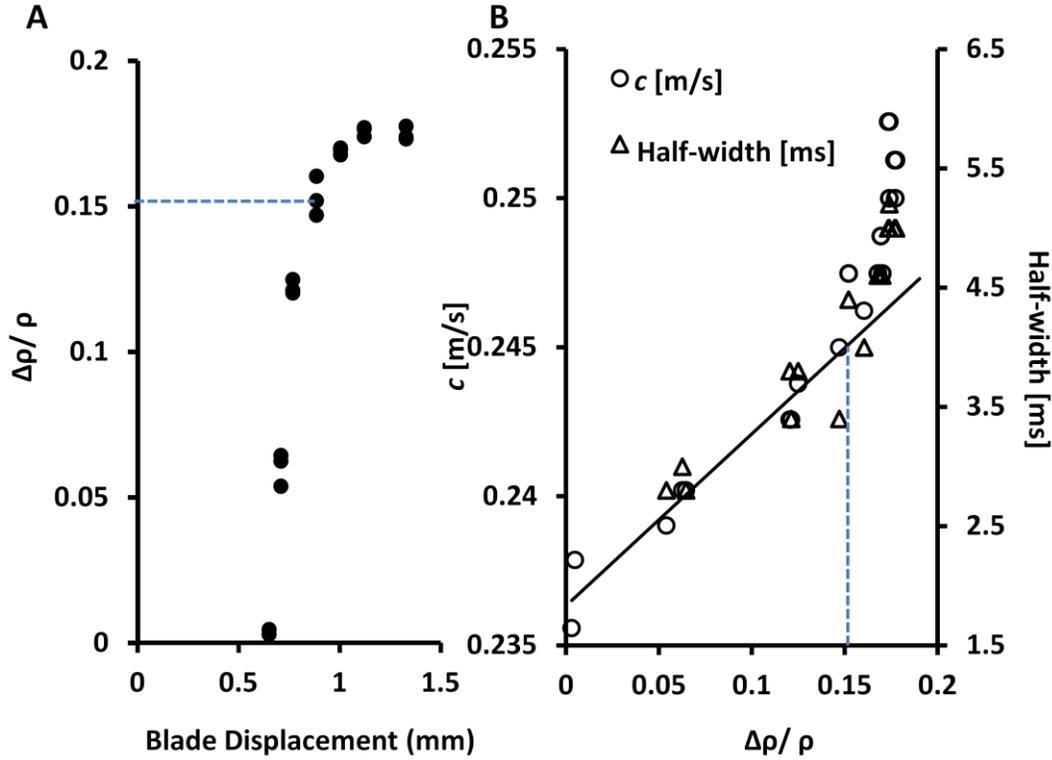

*Figure 2 (Color Online) Threshold, Saturation, Velocity and Pulse Width. (a) Relative compression $\left(\frac{\Delta\rho}{\rho_0}\right)$ extracted from fig. 1a using the quasi-static response curve of fig 1b, as a function of excitations strength (blade's displacement amplitude in mm) (b). The velocity as obtained from the time of arrival of the peak amplitude is plotted with respect to relative compression amplitude $\left(\frac{\Delta\rho}{\rho_0}\right)$. (a,b) The limit of linear dependence on amplitude is indicated by dotted lines.*

Furthermore a first order approximation for the relation between c and $\Delta\rho/\rho_0$ can be written as [24,25];

$$c = c_0 \left[1 + \frac{1}{2}\frac{B}{A}(\Delta\rho/\rho_0)\right] \tag{3}$$

which can be directly compared to the observed dependence of velocity on relative compression ($\Delta\rho/\rho_0$) in Fig. 2b. From the y-intercept of the linear fit $c_0 = 0.236 m/s$ can be directly extracted as the velocity for infinitesimal amplitude (linear limit). Further, the slope of the fit allows the determination of B/A (eq.3) to be 0.5 for our system and using $\Gamma = \frac{1}{2}\frac{B}{A} + 1$ we get $\Gamma = 1.25$ which indeed shows that the curvature of the adiabatic state diagram is positive locally. *However this is in stark contrast with the negative curvature* $\left(\frac{\partial^2 a}{\partial \pi^2}\right)_T$ *of the isothermal state diagram (Fig.2) at the given equilibrium state* ($\pi = 4.3 \frac{mN}{m}$ and $T = 293.15K$) *and hence announces the failure of a quasi-static approximation*. In fact, we believe that the discrepancy between the isothermal and adiabatic curvature is closely related to existence of the observed threshold for excitation: A stable nonlinear wave-front exists only for $\left(\frac{\partial^2 a}{\partial \pi^2}\right)_S > 0$ (see below). When the interface is prepared, however, to exhibit a negative curvature in it's equilibrium state $\left(\frac{\partial^2 a}{\partial \pi^2}\right)_T$, only excitations, which provide sufficient power to transfer the state of the interface from a negative $\left(\frac{\partial^2 a}{\partial \pi^2}\right)_T < 0$ (in the quasi-static limit) to a positive $\left(\frac{\partial^2 a}{\partial \pi^2}\right)_S > 0$ (in the adiabatic limit) while decoupling the interface from the bulk, will result in the formation of stable nonlinear pulses (Fig.3).

**Shocks near Phase Transition – Saturation of Amplitude and Splitting**

Note that a blade displacement of 1.3mm with the rise time of ~5ms gives a maximum particle velocity of ~0.26m/s which is comparable to the velocity of sound in the lipid monolayer in the given state. Hence, the observed pulses can be treated as shock waves and we can learn from classical shock theory. Indeed, in shock compression science $\left(\frac{\partial^2 v}{\partial P^2}\right)_S > 0$ is a necessary condition for stable shocks, which is usually satisfied as in most cases the compressibility ($\sim -\frac{\partial v}{\partial P}$) decreases with pressure. However the exception $\left(\frac{\partial^2 v}{\partial P^2}\right)_S < 0$ can occur near a phase transition or critical points (fig. 3) [18,26–28].

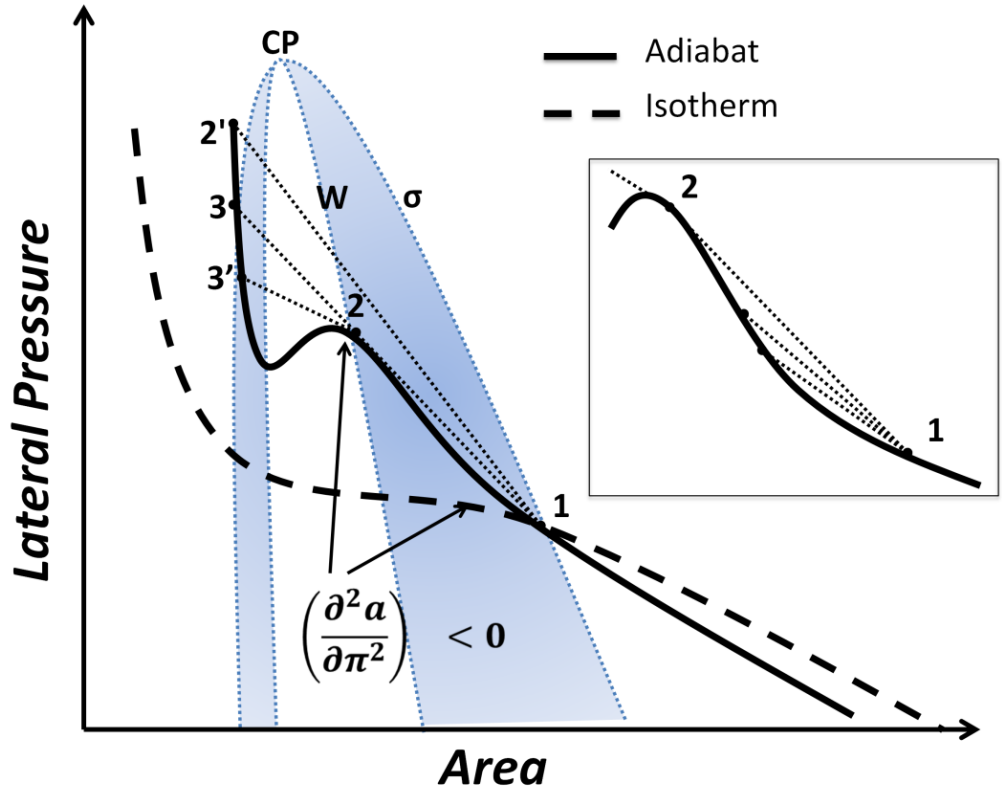

**Figure 3 (Color Online) Representation of the phenomenon on $(\pi \to a)$ State Diagrams.** *Isothermal (dashed) and (shock) adiabatic (solid) state changes are shown. Initially the system is at **1** (equilibrium) prepared right at the equlibrium phase boundary (**σ**). Starting from this point, the non-equilibrium state diagram during a pulse can proceed along various paths determined by the blade velocity. Only if the pulse is excited with enough power to move along a paths in the diagram of $\left(\frac{\partial^2 a}{\partial \pi^2}\right) > 0$ (see text), a solitary wave appears. When observed at a fixed distance (fig.1 and fig.2) the different amplitudes lie on the non-equilibrium adiabat as represented by the solid curve between **1** and **2** with $\left(\frac{\partial^2 a}{\partial \pi^2}\right) > 0$. Since the system was prepared at the equilibrium phase boundary these adiabats extend into the meta-stable (shaded) region, given the positive curvature. For fig. 1 and 2 the end states corresponding to different amplitudes are shown clearly in the inset. The slope of the straight dotted lines is directly related to the shock velocity and represents the jump condition (Rayleigh line) during the shocks. The observed splitting of the pulse (fig. 1 and 4) represents a discontinuity in velocity and hence a discontinuity in the slope of the Rayleigh lines (compare $(\mathbf{1} \to \mathbf{2})$ vs $(\mathbf{2} \to \mathbf{3'})$). This also indicates crossing over the spinodal condition (Wilson line, **W**) or adiabatic*

*phase transition. If we follow a pulse in fig. 4, it's amplitude and hence velocity decreases till the pulse eventually splits at a critical condition represented by ($\mathbf{1 \to 2 \to 3}$). Prior to splitting (closer to the source) the pulse strength may be sufficient to induce a complete phase transition ($\mathbf{1 \to 2'}$). Figure is not to scale. CP indicates the critical point.*

The consequences of a discontinuous and/or negative $\left(\frac{\partial^2 v}{\partial P^2}\right)_S$ in the phase transition region for the stability of a shock-front have been investigated in detail, both theoretically as well as experimentally [18,27,29,30]. One direct observable consequence of $\left(\frac{\partial^2 v}{\partial P^2}\right)_S < 0$, which can occur at phase transition, is the splitting of a compressive shock into a non-dispersive forerunner wave and a slower more dispersive condensation wave near a phase boundary [27]. The emergence of a second wave-front can in fact already be seen in fig. 1a at maximum excitation and indicates a discontinuity within the adiabatic state diagram.

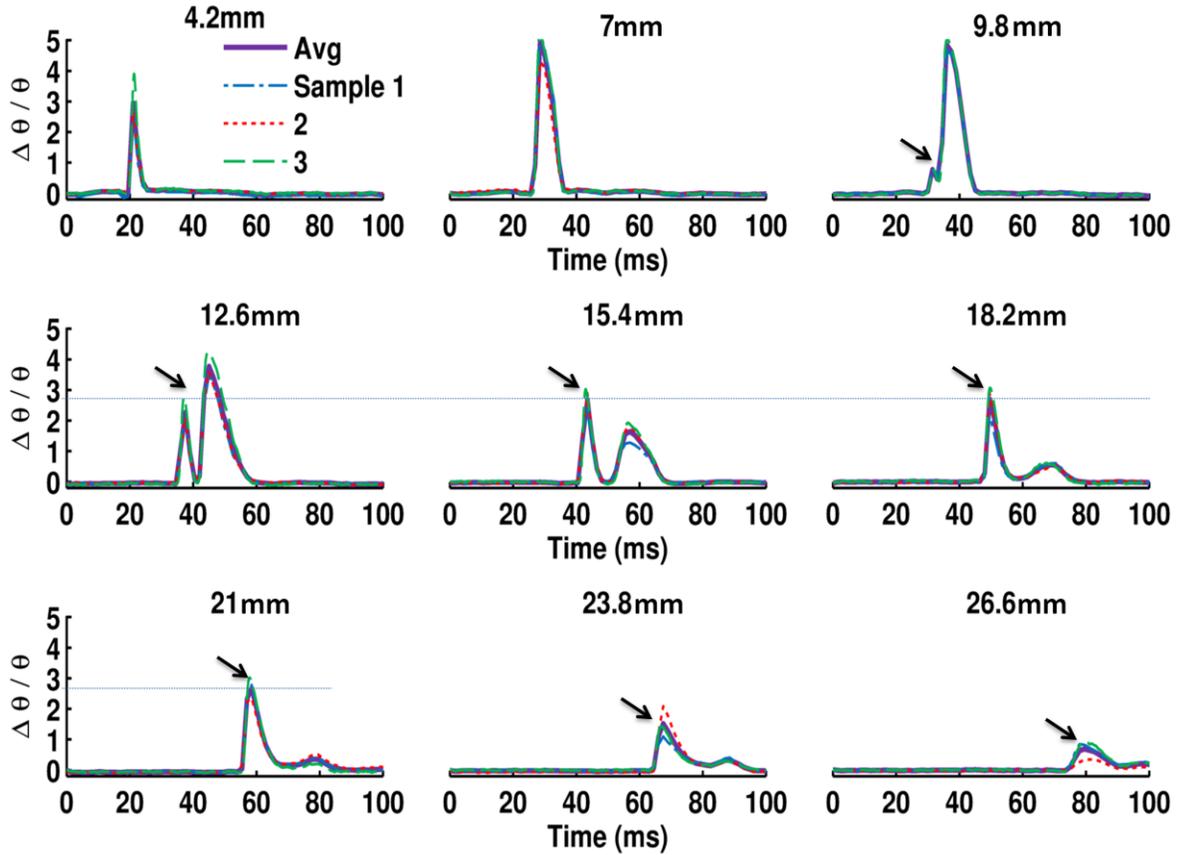

*Figure 4 (Color Online) Distance dependence, evolution and splitting. Pulse shapes are plotted at various distances from the excitation blade (indicated on each plot in mm) for a fixed excitation displacement of 1.3mm and equilibrium state (fig. 1b). After initial solitary propagation up to 7mm, the front runner wave begin to emerge at 9.8mm (indicated by the arrow) and is completely evolved at 12.6mm, while the residual wave that follows shows strong dispersion and loss of amplitude. The frontrunner wave however continues to propagate with stable amplitude and pulse shape for approximately another 1cm up to 21mm (dotted line shows stable amplitude), at which point it starts to disappear as well. To emphasize the repeatability of these experiments pulses from three different experiments and their average is plotted.*

In Fig. 4 the splitting process of the waves – indeed a very rare phenomenon in shock science - is observed in detail. As the pulse shape evolves, it reaches a maximum amplitude at a distance 7mm followed by a decay, which as seen here can result in splitting beginning at 9.8mm. Post splitting, the forerunner wave grows at the cost of the slower more dispersive wave and propagates much further. This along with the fact that the splitting is observed only during decay is consistent with similar observations for vapor-compression shocks in other systems near phase transitions [27]. In order to compare their interpretation with our data it is helpful to assume that the solitary wave profile (Fig. 1 and 4) can be decomposed into a compression wave-front, discussed below, followed by an expansion wave-front both related via the continuity condition[1] (note: in contrast to a compression wave-front, the expansion wave-front is stable for $\left(\frac{\partial^2 a}{\partial \pi^2}\right)_S < 0$ [10].) As the excitation strength is increased beyond threshold, the interface first remains in a regime of positive curvature (hence the increase in $c$ shown in Fig. 2b). We likely cross the metastable regime of the phase transition (see the path in Fig. 3) before the termination of the path at the spinodal, **W** (wave-front (**1** → **2**) in Fig.1 and Fig. 3) or if near the source (**1** → **2′**) which then decays (Fig. 3 and Fig. 4). Eventually the wave becomes instable and splits ((**1** → **2** → **3**) in Fig.3) indicating that a new regime of the adiabatic diagram is entered. Thus instead of simple adiabatic heating during compression, the large amplitude causes nucleation and a phase transition beyond point **2** in Fig.3 (retrograde behavior [27]). Due to the resulting discontinuity at the spinodal line, the "combined" wave (**1** → **2** → **3**) splits into a forerunner shock (**1** → **2**) that propagates – in our case - with a velocity of the liquid-expanded state and a condensation wave (**2** → **3′**), propagating at a slower velocity determined by the properties of the coexistence region of the lipid interface. The interpretation of a pulse-induced nucleation is consistent with the simultaneous observations of (i) a saturation of amplitude, (ii) an abrupt change in curvature and (iii) broadening of pulse shape (increased dispersion), observed in Fig. 1 and 2.

---

[1] In order to follow the lines of Thomson's work [26–28] we imagine our biphasic [6] shock consisting of a compression shock followed by one of rarefaction. In this case shocks begin to "interact" and can for instance weaken each other thereby supporting splitting [27]. However the stability of the rarefaction tail and its interaction with the front will require further analysis and experiments, which are outside the scope of this work and will be treated elsewhere.

It is most likely that similar waves were observed during electrically induced critical de-mixing of multicomponent lipid monolayers by McConell et. al [31,32] and were believed to be shock waves. In these experiments, nucleation resulting from an electrical impulse that propagated as condensed domains, dispersed and dissolved rapidly. Although this non-equilibrium phenomenon was mentioned only qualitatively, these experiments indicate that similar phenomenon can also be observed near complex phase boundaries in multi component systems and can be excited electrically at lipid interfaces.

Nonlinear behavior in terms of a relation between amplitude and velocity has been treated theoretically in the context of models for soliton propagation in isolated lipid bilayers [12,33] and non-equilibrium phase transition in liquid crystals [34]. However, the dispersion relation intrinsic to such models depends not only on state but also on boundary conditions and/or geometry and therefore can be very different, even qualitatively, between a lipid monolayer at the air/water interface and a biological membrane with all its structural complexities. For example, by allowing a quadratic negative nonlinearity in pressure and assuming an adhoc positive linear dispersion, postulated based on measurements of phase velocity in lipid vesicles, Heimburg and Jackson derived a nonlinear wave equation to predict a decrease in velocity and pulse width with increasing amplitude for solitons in lipid membranes [12,35]. This is in complete contrast with the current observations in lipid monolayers where the width and velocity increase with amplitude indicating positive nonlinearity B/A and a highly nonlinear dispersion [36], which is also consistent with the phase transition [37].

We stress, that the line of argument above, relating the curvature of adiabatic state diagrams to the stability and instability of propagating shocks, is universal. This means it will not only hold for mono- and bilayers, but also for an interface in a living system with all its structural complexities. It will be exciting to see experiments on the state diagrams of "living interfaces" and the propagation of nonlinear pulses within them.

An intriguing example of a negative curvature $\left(\frac{\partial^2 a}{\partial \pi^2}\right)_S < 0$ in living systems may be found in nerves near temperature induced reversible blocks for pulse propagation (Fig.5). When approximating the dimensionless curvature $\Gamma = \frac{c^4}{2a^3}\left(\frac{\partial^2 a}{\partial \pi^2}\right)_S$ using Maxwell relations of a simple thermodynamic system[2], an independent thermodynamic equation relating $\frac{B}{A}$ (and hence $\Gamma$) to experimentally accessible variables can be written as [9,38] ($\alpha_T$ being the isothermal expansion and $\tilde{c}_p$ the heat capacity at constant pressure);

$$2(\Gamma - 1) = \frac{B}{A} = 2c_0[\rho_{0A}\left(\frac{\partial c}{\partial \pi}\right)_T + \frac{T\alpha_T}{\rho_{0A}\tilde{c}_p}\left(\frac{\partial c}{\partial T}\right)_\pi] \qquad (4)$$

---

[2] Here we assume a mechanically system, i.e. the state is characterized by $a$, $\pi$ and $T$ only. However, further couplings (Maxwell relations) may play an important role, e.g. electrical (U-q), thermal (E-T) or chemically (µ-N) couplings. These would alter the presented relation.

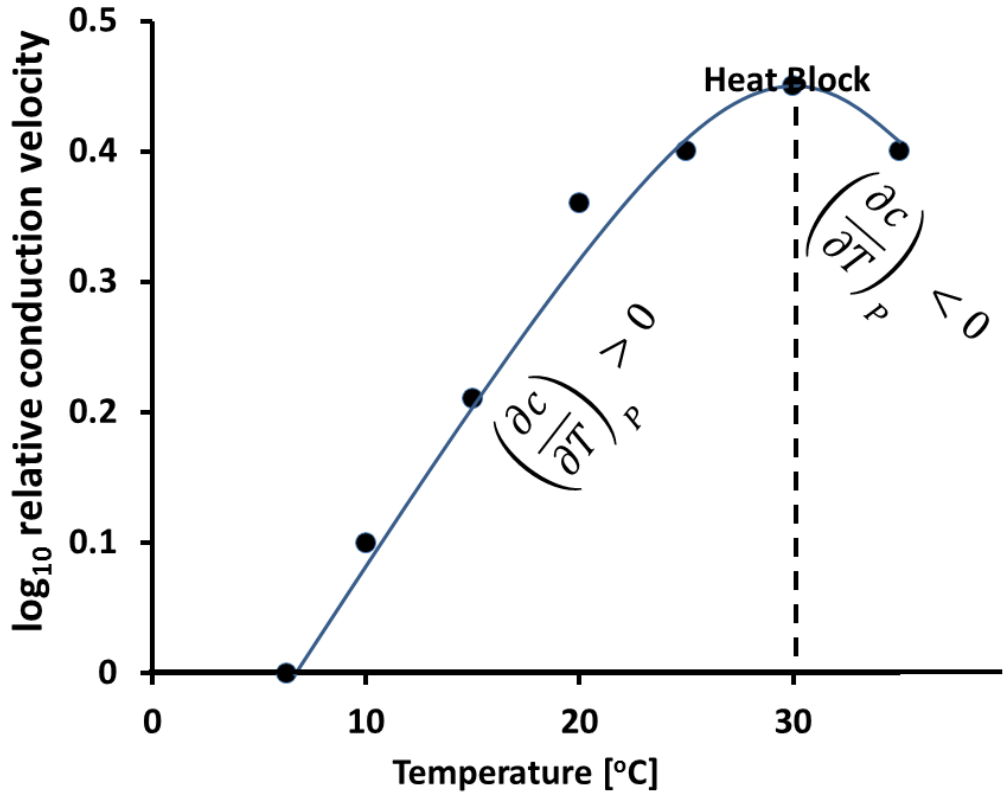

*Figure 5. Heat block of the nerve impulse in squid giant axon (data adapted from [39]). All known action potentials, including those in human, animals or even algae, exhibit a heat block [40]. The temperature varies and can be adapted by changing growth conditions [41] . At the heat block a transition from $\left(\frac{\partial c}{\partial T}\right)_P > 0$ to $\left(\frac{\partial c}{\partial T}\right)_P < 0$ takes place. This condition leads to an instability of the shock wave excited near a phase transition.*

For a lipid monolayer all the variables are experimentally accessible except $\rho_{0A}$, which in fact can be calculated using eq. 4, allowing subsequent calculation of $\Delta\pi$ using eq. 2 (see appendix A). In order to estimate the curvature $\Gamma$, for a nerve fiber, we use the experimentally available [39,42] relations for the bulk pressure and temperature dependence of $c$ , i.e. $\left(\frac{\partial c}{\partial P}\right)_T$ and $\left(\frac{\partial c}{\partial T}\right)_P$ . While experiments demonstrate $\left(\frac{\partial c}{\partial P}\right)_T$ is usually negative[3] [42,43], the quantity $\left(\frac{\partial c}{\partial T}\right)_P$ starts of positive but gets increasingly negative as we approach the temperature corresponding to heat block [39,44]. Indeed negative values of $\left(\frac{\partial c}{\partial P}\right)_T$ and $\left(\frac{\partial c}{\partial T}\right)_P$ imply a negative value of B/A and – if less than -2 – a negative Γ as well, which would lead to the above mentioned cessation of the shock wave.

In conclusion we have demonstrated the application of shock compression science at a soft interface near phase transition and its implications for biological systems, especially nerves. In particular we extracted the curvature $\left(\frac{\partial^2 a}{\partial \pi^2}\right)_S$ of the adiabatic state diagram from solitary waves observed in lipid monolayer and tied it to the observations of excitation threshold, amplitude saturation and stability of solitary waves (against splitting). Since this is the first attempt to apply shock compression science to soft interfaces, several questions remain open and predictions to be tested: For example our approach predicts the culmination of metastable regimes into a critical point near heat-block in nerves [11,45]. Furthermore the saturation in amplitude, which in our case originates from crossing over the spinodal condition, should correspond to a "dynamic" phase transition in nerve pulse propagation, which remains to be observed [46]. Another aspect arises from the fact, that opto-mechancial coupling deserves some attention, as the calibration is done for the isothermal case [20]. Obviously, the interpretation of the experiments in the present work has been intentionally oversimplified as for a more quantitative description, further analysis, both theoretical and experimental is required. New insights will arise from challenging the quasi-2D nature of the propagation, as the role of thickness of the hydration layer as well its interaction with the bulk or other interfaces nearby needs to be investigated. In addition, the coupling of further thermodynamic variables (lipid dipole, bulk pH, charge) and the role of boundary conditions will be important for a deeper understanding of these pulses in biology[4]. This would also lead to a better understanding of dissipation in our system, since even though we find regimes where the pulse can cover significant distance with constant amplitude and width (fig. 4), dissipation of our solitary waves remains to be a crucial unresolved issue when comparing them to nerve impulses. Finally the interaction of propagating shock waves with the complex chemistry of biological interfaces (e.g. enzymes) might lead to completely new phenomena that can now be studied systematically [47]. We believe this work opens new doors for the physics community to contribute to life sciences, in particular we imagine, to the understanding of inter- and intra-cellular communication by combining nonlinear acoustics and the physics of critical phenomenon at interfaces.

---

[3] precise measurements of $\left(\frac{\partial c}{\partial P}\right)_T$ near the heat block to clarify the sign are currently not available. It seems plausible, however, to assume a decrease in c with pressure even at temperatures near heat block.

**APPENDIX A**

It is assumed that the lipid monolayer along with a few hydration layers form the propagation medium that is adiabatically decoupled from the bulk. Hence $\rho_{0A}$ with its dimensions of kg/m² essentially represents the mass of this medium projected on a 2D interface. In order to calculate $\rho_{0A}$ we employ the following independent thermodynamic relation for B/A;

$$\frac{B}{A} = 2c_0[\rho_{0A}\left(\frac{\partial c}{\partial \pi}\right)_T + \frac{T\alpha_T}{\rho_{0A}C_p}\left(\frac{\partial c}{\partial T}\right)_\pi] \tag{5}$$

---

[4] The condition $\frac{\partial^2 v}{\partial P^2} > 0$ results from $\Delta S > 0$ in a purely mechanical system [18] which implies entropy is a function of only the volume. However in general entropy is a multi-dimensional function of all the extensive variables of the interface.

While B/A is known from the experimental amplitude-velocity relation, all the other parameters are accessible except $\rho_{0A}$. For example, the state $\pi = 4.3 mN/m$ and T=293.15K corresponds to the edge of the transition region and we have previously shown [5] that velocity decreases from 0.4m/s to 0.2m/s on increasing pressure by 1mN giving $\frac{\Delta c}{\Delta \pi} = -0.2 \, m/mNs$ at constant T. An increase in temperature at constant pressure on the other hand, moves the system away from the transition region resulting in an increase in velocity, here found experimentally to be $\frac{\Delta c_0}{\Delta T} = 0.009 m/Ks$ at constant π=4.3mN/m. Finally, the phenomenon takes place across a LE-LC phase transition, allowing the approximation $\frac{T\alpha_T}{\rho_0 C_p} = \frac{dT_{tr}}{d\pi_{tr}}$ [48] (the subscript indicates that the pressure and temperature are taken at the transition). Plugging in the values gives $\rho_{0A} = 0.012 kg/m^2$ completing the equation of state Eq. 1, defined locally for $\pi = 4.3 mN/m$ T=293.15K. As a result, Δπ can be calculated and is found to be 0.12 mN/m, which is in good agreement with the previously reported values of pressure pulses in the transition region [5]. Note that the obtained value of $\rho_{0A}$ is a gross overestimate of the actual value as it also accounts for charge/dipole effects that have been completely ignored in Eq.1 and Eq.5.

**Acknowledgments:** We thank Dr. Konrad Kauffman (Göttingen) who introduced us to the thermodynamic origin of nerve pulse propagation and its theoretical explanation. We would also like to thank him for numerous seminars and discussions. We would like to thank Dr. Christian Fillafer and Prof. Glynn Holt for helpful discussions and critical reading of the manuscript. Financial support by BU-ENG-ME is acknowledged. MFS appreciates funds for guest professorship from the German research foundation (DFG), SHENC-research unit FOR 1543.